\documentclass{moriond}

\def\be{\begin{equation}}
\def\ee{\end{equation}}
\def\bea{\begin{eqnarray}}
\def\eea{\end{eqnarray}}

\newcommand{\Photo}{\includegraphics[height=35mm]{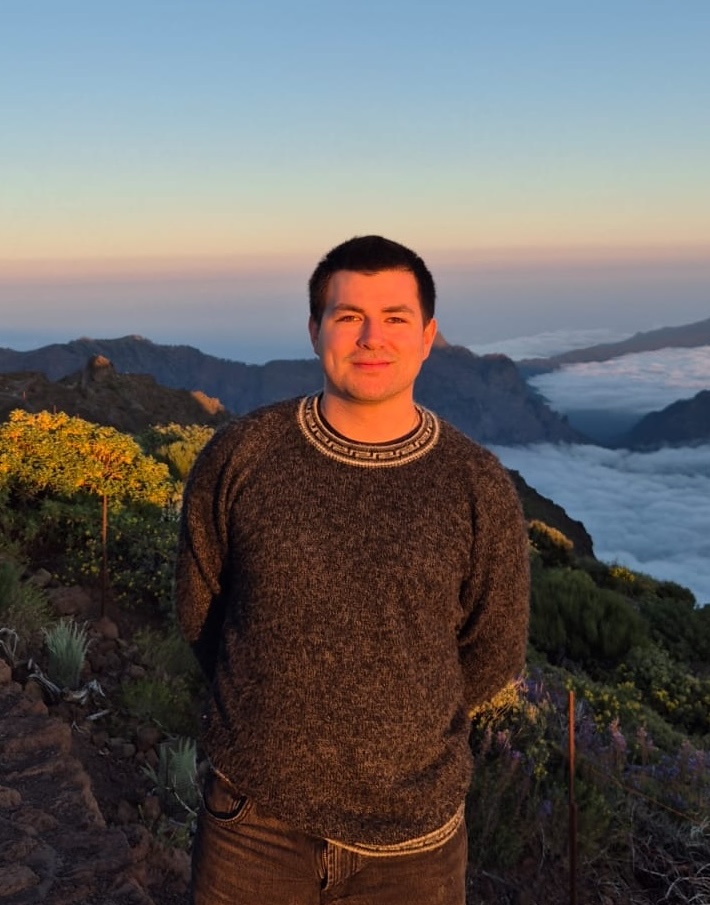}}

\begin{document}
\vspace*{4cm}
\title{Enabling real-time multi-messenger follow-up of transient events with Astro-COLIBRI}

\author{Bernardo Cornejo Avila, S. Bisero, M. Costa, A. Ciric, I. Jaroschewski, W. Kiendr\'{e}b\'{e}ogo, and F. Schussler}

\address{IRFU, CEA, Universit\'{e} Paris-Saclay,\\F-91191 Gif-sur-Yvette, France}

\maketitle\abstracts{
Time-domain astrophysics is a rapidly growing field focused on the study of transient phenomena such as Gamma-Ray Bursts (GRBs), Fast Radio Bursts (FRBs), supernovae, novae, and AGN flares. Their characterization increasingly relies on a multi-messenger and multi-wavelength approach, combining gravitational waves, high-energy neutrinos, and electromagnetic observations across the spectrum. Such a coordinated strategy requires efficient information sharing and thus tools capable of rapidly compiling and contextualizing key data for each new event. We present Astro-COLIBRI, a well-established platform designed to meet this challenge.
Astro-COLIBRI combines a public RESTful API, real-time databases, and a cloud-based alert system. It continuously listens to multiple alert streams, applies user-defined filters, and places each event in its multi-messenger and multi-wavelength context. Through its user-friendly interfaces, including a web application and mobile apps for iOS and Android, the platform provides clear data visualization as well as concise summaries of key event properties and observing conditions for user-defined locations.}

\section{Introduction}

The detection and characterization of high-energy transient phenomena come with many observational challenges. Rapid responses to alerts from multiple information streams require a high level of communication between different actors. In addition, while the emergence of new cosmic messengers, such as astrophysical neutrinos and gravitational waves, has provided new perspectives, it has also highlighted the importance of coordinated multi-messenger and multi-wavelength observations.

Current strategies rely on wide field-of-view telescopes monitoring the night sky. Once an event is detected, a public alert is distributed through dedicated brokers such as GCN \footnote{https://gcn.nasa.gov/}, TNS \footnote{https://www.wis-tns.org/}, or Fink \footnote{https://fink-broker.org/}, as well as through internal communication channels. These alerts are received by pointed observatories, which offer better resolution, sensitivity to different energy ranges, or access to other messengers. The observatories then decide whether to perform follow-up observations on the alert position. Some responses are automated, such as the one developed by the H.E.S.S. collaboration~\cite{hess_transients}, while others rely on manual decisions based on the available information about the source, often spread across many sites. 

In this context, Astro-COLIBRI~\cite{astrocolibri} was developed as a downstream tool designed to facilitate transient follow-up observations. It is a platform that collects, filters and contextualizes transient alerts from multiple sources. The system continuously listens to real-time alert streams, applies user-defined filters, and places each event within its multi-wavelength and multi-messenger context. Through a user-friendly interface available via web \footnote{https://astro-colibri.com/} and mobile applications, Astro-COLIBRI provides concise summaries of key event properties, observing conditions for user-defined locations, and real-time notifications, allowing both professional and amateur astronomers to perform rapid and coordinated follow-up observations of sources of interest.

\begin{figure}[ht]
\centering
\includegraphics[width=0.6\linewidth]{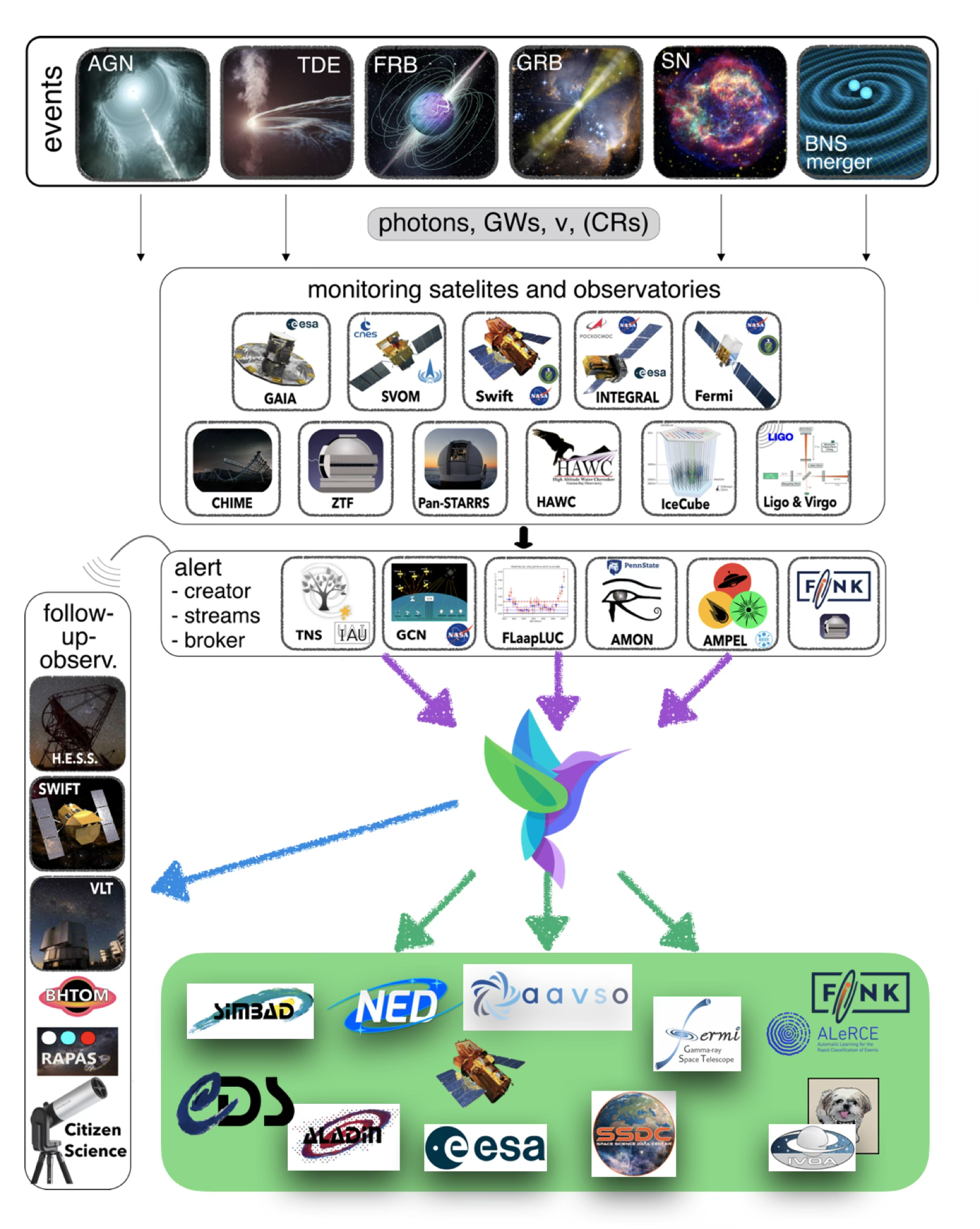}
\caption{Diagram of the multi-messenger transient follow-up strategy, highlighting Astro-COLIBRI as a downstream coordination platform.}
\label{fig:ac_diagram}
\end{figure}

\section{Astro-COLIBRI}

\subsection{Graphical user interfaces and main functionalities}

\begin{figure}[ht]
\centering
\includegraphics[width=0.8\linewidth]{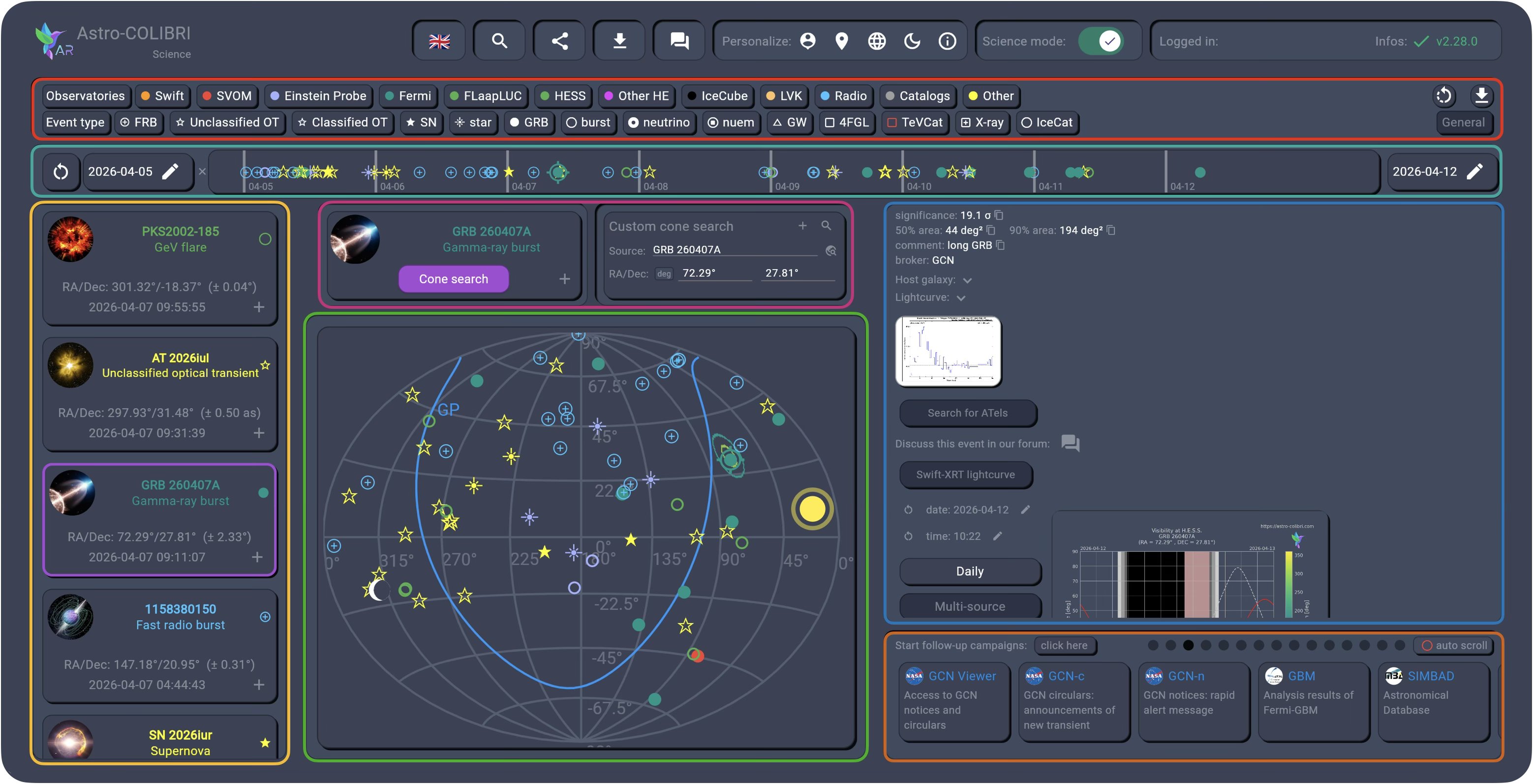}
\caption{Astro-COLIBRI web front-end divided into its panel structure. \textit{Red}: filter panel, \textit{Light blue}: timeline panel, \textit{Yellow}: card tile panel, \textit{Violet}: cone search panel, \textit{Green}: sky map panel, \textit{Blue}: information panel, \textit{Orange}: external links panel.}
\label{fig:ac_frontend}
\end{figure}

Astro-COLIBRI provides two easy-to-use graphical user interfaces: a web platform and a mobile application. Both versions share the same functionalities, including push notifications for new and updated events. The interface is organized into panels, highlighted in different colors in Figure \ref{fig:ac_frontend}, designed to provide intuitive access to different features:

\begin{itemize}
    \item \textbf{Timeline panel} (\textit{light blue}): Latest transients are arranged in chronological order, highlighting the dates within the selected time period. By default, this panel shows events from the last 7 days, but the time range can be customized by the user. 
    \item \textbf{Event tiles panel} (\textit{yellow}): Events are displayed using card tiles in the left panel, allowing users to navigate and select sources. An event can be added to the watchlist, remaining accessible even after the selected time period has passed. If an event is missing from the database, users can manually create one to save a specific location in the sky. 
    \item \textbf{Sky map panel} (\textit{green}): Events are placed according to their spatial coordinates on a central sky map. The selected source is highlighted, and users can freely zoom and pan the map. Events with a large localization uncertainty have their 68\% and 95\% contours displayed. The coordinate system and projection can also be customized.
    \item \textbf{Cone search panel} (\textit{violet}): The cone search button allows users to display transients and high-energy archival events within a 10-degree radius from the selected source. It can also be performed at custom sky locations.
    \item \textbf{Information panel} (\textit{blue}): The right section provides key information about the selected event, helping users decide whether follow-up observations are possible and necessary. In addition to parameter values, public plots are displayed (e.g. Fermi-GBM light curves in the case of Fermi triggered GRBs), and event-dependent plots are produced to place them into their archival context (e.g. detection magnitude vs redshift for supernovae). Other information, such as photometry for optical transients, is also included. Additional on-demand requests are also available, including a tiling scheduler for poorly localized events supported by the tilepy~\cite{tilepy} software or light curve generation from public data for optical sources. Finally, visibility curves from the user location or preferred observatory are displayed, with options for multi-observatory, multi-night, and multi-source configurations. 
    \item \textbf{External links panel} (\textit{orange}): At the bottom right there is a set of URLs pointing to related information pages, such as official detection reports or other astrophysical tools. This panel also holds observatory-specific follow-up buttons, allowing users to submit target of opportunity observation requests using each observatory's procedure (user-restricted).
    \item \textbf{Filter panel} (\textit{red}): The top panel contains observatory and event-type filters. Users can activate or deactivate filters to display only events of interest. Most filters provide additional sub-filters, accessible on long-press, including magnitude for optical sources, distance, or localization uncertainty, among others.
    \item \textbf{Top menu buttons}: At the top of the page, a set of buttons allows users to search for sources, change the language, share events, customize their location, create an account, discuss events in the forum, and download selected events.
\end{itemize}

The described structure corresponds to the web platform. The mobile application version follows the same logic and functionality, adapted to different screen sizes and resolutions. All features are supported by a public RESTful API, which can also be accessed directly through a set of endpoints for integration into user scripts. The API documentation page \footnote{https://astro-colibri.science/apidoc} describes how to use these endpoints and provides detailed examples in the form of notebooks.

\subsection{The platform architecture}

Astro-COLIBRI is built around a public RESTful API, acting as a bridge between two real-time databases and the user-facing clients used to interact with the most recent transient alerts. This dual-database system enables real-time updates while maintaining a long-term archive, which currently contains more than 100,000 events since 2016. A continuously running listener handles the different alert streams, places each alert into its context, and communicates new events and updates to users through push notifications distributed by a cloud-based alert system. The API is then responsible for plot generation, database storage, and handling user requests from the front-end.

\section{Outlook}

Transient astrophysics is a rapidly evolving field driven by new facilities, improved instrumentation, and increasingly efficient communication strategies. In this dynamic context, Astro-COLIBRI has established itself as a valuable tool dedicated to facilitating follow-up observations. Through user-friendly interfaces and innovative features, it provides researchers and amateur astronomers with an all-in-one platform for newly detected sources, reducing response times and increasing follow-up rates, thereby enabling further progress in the field.

An active and accessible development team, together with an engaged user community, ensures that the platform remains up to date with new methods and developments. Continuous improvements and new implementations, driven by community feedback, are among the strengths of Astro-COLIBRI, allowing it to remain a key tool for transient follow-up in the years to come.

\section*{Acknowledgments}

The authors acknowledge the support of the French Agence Nationale de la Recherche (ANR) via the MOTS project (reference ANR-22-CE31-0012). This work is also supported by the European Union's Horizon Europe Research and Innovation program under the ACME project (grant agreement n. 101131928).

\section*{References}
\bibliography{moriond}

%%% manually generated bibliography
%\begin{thebibliography}{99}
%\bibitem{ja}C Jarlskog in {\em CP Violation}, ed. C Jarlskog
%(World Scientific, Singapore, 1988).
%\bibitem{ma}L. Maiani, \Journal{\PLB}{62}{183}{1976}.
%\bibitem{bu}J.D. Bjorken and I. Dunietz, \Journal{\PRD}{36}{2109}{1987}.
%\bibitem{bd}C.D. Buchanan {\it et al}, \Journal{\PRD}{45}{4088}{1992}.
%\end{thebibliography}

\end{document}